\documentclass[aps,pra,twocolumn,amsmath,nofootinbib]{revtex4-1}
\usepackage{epsfig}
\usepackage[colorlinks,linkcolor=blue,anchorcolor=blue,
            citecolor=blue,urlcolor=blue,
            breaklinks=true]{hyperref}
\usepackage{graphics}
\usepackage{color}
\usepackage{graphicx}% Include figure files
\usepackage{dcolumn}% Align table columns on decimal %point
\usepackage{bm}% Bold math
\usepackage{cases}
\usepackage{amsfonts}
\usepackage{appendix}
\usepackage{ulem}

\newcommand{\tb}{\textbf}

\begin{document}
\author{Yong-Long Wang$^{1,2}$}
 \email{Email: wangyonglong@lyu.edu.cn}
\author{Hao Zhao$^{1, 2}$}
\author{Hua Jiang$^{1}$}
%\author{Cheng-Zhi Ye$^{1}$}
\author{Hui Liu$^{2}$}
\email{liuhui@nju.edu.cn}
\author{Yan-Feng Chen$^{3}$}
\email{Email: yfchen@nju.edu.cn}
%\author{Hong-Shi Zong$^{2,4,5}$}
%\email{Email: zonghs@nju.edu.cn}
\address{$^{1}$ School of Physics and Electronic Engineering, Linyi University, Linyi, 276000, China}
\address{$^{2}$ Department of Physics, Nanjing University, Nanjing 210093, China}
\address{$^{3}$ National Laboratory of Solid State Microstructures, Department of Materials Science and Engineering, Nanjing University, Nanjing, 210093, China}
%\address{$^{4}$ Department of Physics, Anhui Normal University, Wuhu, Anhui 241000, China}
%\address{$^{5}$ Nanjing Institute of Proton Source Technology, Nanjing 210046, China}

\title{Geometry-induced Monopole Magnetic Field and Quantum Spin Hall Effect}
\begin{abstract}
The geometric effects of two-dimensional curved systems have been an interesting topic for a long time. A M\"{o}bius surface is specifically considered. For a relativistic particle confined to the nontrivial surface, we give the effective Dirac equation in the thin-layer quantization formalism, and we find a geometric gauge potential that results from the rotation transformation of the local frame moving on M\"obius strip, and an effective mass that is from the rescaling transformation. Intriguingly, the geometric gauge potential can play a role of monopole magnetic field for the particles with spin, and which can produce quantum spin Hall effects. As potential applications, effective monopole magnetic fields and spin Hall phenomena can be generated and manipulated by designing the geometries and topologies of two-dimensional nanodevices.
\bigskip

% Electron states at surfaces and interfaces, 73.20.-r;
% Electronic transport phenomena in thin film, 73.50.-h;
% Quantum mechanics, 03.65.-w;
% Geometry, differential geometry, and topology, 02.40.-k;
\end{abstract}
\maketitle

\section{Introduction}\label{1}
With the rapid development of artificial microstructure technology, theoretical and experimental physicists are always trying to discover novel phenomena of Hall physics in the thin films with complex geometries. Those investigations boost the interests in the effective quantum dynamics of curved surface. A much suitable scheme, the thin-layer quantization approach was primitively employed to study the geometric quantum effects of curved surface by introducing a confining potential~\cite{Jensen1971Quantum, Costa1981Quantum}, and then the approach was generalised to low-dimensional manifolds embedded in high-dimensional manifolds~\cite{Jaffe2003Quantum}. In order to eliminate the ambiguity of calculation order, the quantization approach was clearly regularized in a fundamental formalism~\cite{Wang2016Quantum}, in which the geometric effects mainly manifest as a scalar geometric potential~\cite{Costa1981Quantum}, a geometric momentum~\cite{Liu2011Geometric, Wang2017Geometric}, a geometric orbital angular momentum~\cite{Wang2017Geometric}, and a geometric gauge potential~\cite{Jaffe2003Quantum, Wang2018Geometric}. The scalar geometric potential has been proved that can construct a topological band structure for periodically minimal surfaces~\cite{Aoki2001Electronic}, can generate bound states for spirally rolled-up nanotubes~\cite{Ortix2010Effect}, can eliminate the reflection for bent waveguides~\cite{Campo2014Bent}, can provide the transmission gaps for periodically corrugated thin layers~\cite{Wang2016Transmission, Wang2019The} and so on. The geometric momentum and the geometric angular momentum can additionally contribute~\cite{Lai2019Geometrical} and modify the spin-orbit coupling~\cite{Ortix2014Absence, Wang2017Geometric, Armitage2018Weyl}. As empirical evidences, the scalar geometric potential has been realized by an optical analogue in a topological crystal~\cite{Szameit2010Geometric}, and the geometric momentum was observed to affect the propagation of surface plasmon polaritons on metallic wires~\cite{Schmidt2015Guided}. In other words, the thin-layer quantization formalism is valid for a curved surface system.

As far as we know, the thin-layer quantization formalism has been successfully employed to deduce the effective Schr\"odinger equation~\cite{Jensen1971Quantum, Costa1981Quantum, Costa1982Constraints, Encinosa1998Energy,Encinosa2003Curvature, Gravesen2005Schrodinger, Ferrari2008Shcrodinger, Jensen2009Quantum, Ortix2011Absence, Oliveira2014Quantum} and the effective Pauli equation~\cite{Kosugi2011Pauli, Kenneth2014Braiding, Wang2014Pauli}. Most of the known geometric effects are induced by curvature that is determined by the relationship between the three-dimensional metric tensor and the two-dimensional metric tensor~\cite{Jensen1971Quantum, Costa1981Quantum}, which can be also determined by a diffeomorphism transformation. As a crucial ingredient, the geometric potential in the non-relativistic curved system does not appear in the relativistic case, the effective Dirac equation~\cite{Matsutani1993Quantum, Jensen1993Fermions, Brandt2016Dirac, Liu2020No}. Therefore, for the relativistic particle confined to a curved surface the geometric effects need a further investigation. Due to the spinor being not a representation of a dffeomorphism group, we have to consider not only the diffeomorphism transformations, but also the rotation transformation of dreibein fields that is a local Lorentz rotation~\cite{Parrikar2014Torsion}. In the thin-layer quantization formalism, the effective quantum dynamics is obtained by reducing the normal degree of freedom, and the geometric effects can be induced not only by curvature, but also by torsion~\cite{Wang2018Geometric}. The torsion-induced effects have been demonstrated in a twisted quantum ring~\cite{Taira2010Torsion}, on a M\"{o}bius ladder~\cite{Sun2009Observable, Souza2017Aharonov} and a space curve~\cite{Brandt2015Induced} as a torsion-induced magnetic moment, a torsion-induced Zeeman-like coupling and an anomalous phase shift~\cite{Taira2010Anomalous}, respectively. Therefore, in the torsion-induced gauge structure, the nontrivial properties of geometric magnetic field and the topological properties of torsion Landau levels need further investigations.

The Dirac magnetic monopole is an Abelian monopole that results from the singularities of electromagnetic field~\cite{Dirac1931}. Decades later a non-Abelian magnetic field was generalised for the non-Abelian Yang-Mills gauge field~\cite{Yang1975Concept, Kondo2018Magnetic, Fujikawa2020A}. Although the magnetic monopole is not an active research topic, the related researches are still published from time to time. Theoretically, the magnetic monopole was constructed in the pure Yang-Mills theory~\cite{Yang1975Concept}, in the Georgi-Glashow model~\cite{tHooft1974Magnetic} and in the "complementary" gauge-scalar model~\cite{Kondo2016Gauge}. Experimentally, the magnetic monopole has been proposed to be realised in bilayer graphene~\cite{San2012Non, Uri2020Nanoscale}. Particularly, the discussions of magnetic monopole were triggered by the appearance of SU(2) gauge theory in the classification of four-manifolds~\cite{Donaldson1990The}. In a most general case, the effective quantum dynamics on a two-dimensional manifold or a one-dimensional manifold can be endowed SU(2) gauge structures by reducing two degrees of freedom with a confining potential of SO(3) symmetries~\cite{Brandt2015Induced, Liang2020Effective}. For a specific case, it was proved that the effective dynamics confined to a certain curved surface is endowed U(1) gauge structure by introducing a confining potential with SO(2) symmetries~\cite{Wang2020Geometry}. Mathematically, those Abelian and non-Abelian gauge fields can be induced by the geometries of curved systems that is like the strain-induced pseudomagnetic field~\cite{Chang2014Strain} and the tensor monopole~\cite{Zhu2021Experimental}. As a crucial result in the present paper, an effective SU(2) gauge potential can be induced by the geometries of M\"obius strip for spin, and behaves as a monopole effective magnetic field.

A new topological state of quantum matter, the quantum spin Hall effect is different from the traditional quantum Hall effect without externally applying magnetic fields. It is strikingly that the quantum spin Hall effect was primitively and independently predicted as the effects of spin-orbit interactions~\cite{Kane2005Quantum, Kane2005Z2} and as the result of the presence of a strain gradients~\cite{Zhang2006Quantum}. Subsequently, the new topological phenomenon was experimentally realised in HgTe quantum wells~\cite{Zhang2006QuantumB}. For the quantum Hall effect, the quantized Hall conductances are determined by the quantum Landau levels that are created by the externally applied magnetic fields, while the quantum spin Hall effect is determined by the degenerate quantum Landau levels that is created by the spin-orbit coupling in conventional semiconductors~\cite{Kane2010Colloquium, Qi2011Topological}. Therefore, it is relative hard to realise experimentally. In view of that the geometries of curved systems can play the role of gauge potential for particles with orbital spin~\cite{Olpak2012Dirac, Wang2018Geometric}, therefore there appears an interesting topic to study the relationship between the topologies of quantum Hall state and those of curved systems. As another important result, the quantum spin Hall effects can be induced by the geometry of M\"obius strip.

As previously mentioned, we will deduce the effective Dirac equation for a relativistic particle that is confined to M\"obius strip, and will specifically study the quantum effects that are induced by the nontrivial properties of M\"obius strip. The present paper is organized as follows. In Sec. the geometrical properties of a M\"obius strip are reviewed. In Sec. III, the thin-layer quantization formalism is briefly discussed, and then it is employed to deduce the effective Dirac equation for the relativistic particle confined to M\"obius strip. In Sec. IV, the geometry-induced monopole magnetic field is discussed. In Sec. V, the quantum spin Hall effects induced by geometry are investigated. In Sec. V, the conclusions and discussions are given.

\section{The geometrical properties of M\"{o}bius strip}
For a curved surface $\mathbb{S}^2$, one can adapt a suitable curvilinear coordinate system to describe it as $\tb{r}(q_1, q_2)$, a function of $q_1$ and $q_2$, where $q_1$ and $q_2$ are two tangent coordinate variables of $\mathbb{S}^2$. For a point near $\mathbb{S}^2$, one has to introduce a coordinate variable $q_3$ normal to $\mathbb{S}^2$ to parameterize it as $\tb{R}(q_1, q_2, q_3)=\tb{r}(q_1, q_2)+q_3\tb{n}$, where $\tb{n}$ denotes the basis vector normal to $\mathbb{S}^2$. The presence of $\mathbb{S}^2$ will deform its near space that is denoted as $\Xi\mathbb{S}^2$. The deformation can be described by a diffeomorphism transformation which belongs to $GL(3, \mathbb{R})$. As usual, a relativistic particle can be described by a spinor. Due to without a spinor representation, $GL(3, \mathbb{R})$ can not be employed to describe the action of diffeomorphism transformation on the spinor that is confined to $\mathbb{S}^2$. However, the confined spinor will obey the rotation transformation connecting the local frames of the different points on $\mathbb{S}^2$~\cite{Bertlmann1996Anomalies}, which is a generator of $SO(3)$. In what follows, a M\"obius strip will be considered.

\begin{figure}[htbp]
  \centering
  % Requires \usepackage{graphicx}
  \includegraphics[width=0.36\textwidth]{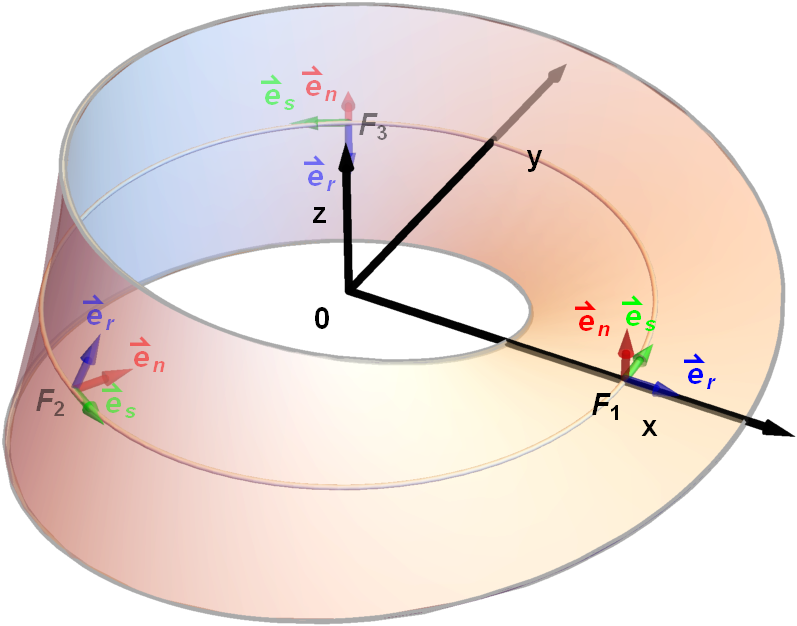}
  \caption{\footnotesize Schematic of a M\"obius strip. $\vec{e}_r$ and $\vec{e}_s$ are two tangent unit basis vectors, and $\vec{e}_n$ is the normal unit basis vector of M\"obius strip. The local frames $F_1$, $F_2$ and $F_3$ are localized at $\theta=0, \frac{4\pi}{3}, \frac{8\pi}{3}$, respectively. }\label{Fig1}
\end{figure}
A M\"obius surface of half-width $w$ with midcircle of radius $R$ and at height $z=0$, which can be parametrized by $\tb{r}(r, \theta)=(r_x, r_y, r_z)$ with
\begin{equation}\label{Mobius}
\begin{split}
& r_x(r, \theta)=[R+r\cos(\frac{\theta}{2})]\cos\theta,\\
& r_y(r, \theta)=[R+r\cos(\frac{\theta}{2})]\sin\theta,\\
& r_z(r, \theta)=r\sin(\frac{\theta}{2}),
\end{split}
\end{equation}
where $r$ and $\theta$ are two variables with $r\in [-w, w]$ and $\theta\in [0, 4\pi]$, $R$ can be taken as a constant. The M\"obius strip is denoted by $\mathbb{M}^2$ and sketched in Fig.~\ref{Fig1}. For convenience, we can adapt an orthogonal frame spanned by two tangent vectors $\tb{e}_{r}$ and $\tb{e}_s$ and a normal vector $\tb{e}_n$. According to Eq.~\eqref{Mobius}, we can obtain the three basis vectors as
\begin{equation}\label{BVector}
\begin{split}
& \tb{e}_r=(\cos\frac{\theta}{2}\cos\theta,\cos\frac{\theta}{2}\sin\theta, \sin\frac{\theta}{2}),\\
& \tb{e}_s=\frac{2}{N}(-(R\sin\theta+\frac{3}{2}r\cos\theta\sin\frac{\theta}{2}+r\sin\frac{\theta}{2}),\\
&\quad\quad\quad R\cos\theta+\frac{1}{4}r\cos\frac{\theta}{2}+\frac{3}{4}r\cos\frac{3\theta}{2}, \frac{1}{2}r\cos\frac{\theta}{2}), \\
& \tb{e}_n=\frac{2}{N}(R\sin\frac{\theta}{2}\cos\theta-r\sin^2\frac{\theta}{2}\sin\theta,\\
& \quad\quad\quad R\sin\frac{\theta}{2}\sin\theta+\frac{1}{2}r(\sin^2\theta+\cos\theta),\\
& \quad\quad\quad -R\cos\frac{\theta}{2}-r\cos^2\frac{\theta}{2}),
\end{split}
\end{equation}
respectively, where $N$ is a normalized constant as
\begin{equation}\label{GExpress1}
N=(4R^2+8Rr\cos\frac{\theta}{2}+2r^2\cos\theta+3r^2)^{\frac{1}{2}}.
\end{equation}
It is straightforward that $(\tb{e}_r, \tb{e}_s, \tb{e}_{n})$ can be obtained from the usual basis vectors $(\tb{e}_x, \tb{e}_y, \tb{e}_z)$ by a rotation transformation ${\rm{U}}_{\mathcal{R}}$ in the following form
\begin{equation}\label{CartToCurv}
\left [
\begin{array}{ccc}
\tb{e}_r\\
\tb{e}_s\\
\tb{e}_n
\end{array}
\right ]={\rm{U}}_{\mathcal{R}}(r, \theta)\left [
\begin{array}{ccc}
\tb{e}_x\\
\tb{e}_y\\
\tb{e}_z
\end{array}
\right ],
\end{equation}
where ${\rm{U}}_{\mathcal{R}}$ can be specifically expressed as
\begin{equation}\label{RotTrans}
{\rm{U}}_{\mathcal{R}}(r, \theta)=\left [
\begin{array}{ccc}
{e_{r}}^{x} & {e_{r}}^{y} & {e_{r}}^{z}\\
{e_{s}}^{x} & {e_{s}}^{y} & {e_{s}}^{z}\\
{e_{n}}^{x} & {e_{n}}^{y} & {e_{n}}^{z}
\end{array}
\right ],
\end{equation}
through the dreibeins ${e_{\alpha}}^{i}$, $(\alpha=r, s, n)$ and $(i=x, y, z)$, which can be written as
\begin{equation}\label{Veilbein}
\begin{split}
& {e_{r}}^{x}=\cos\frac{\theta}{2}\cos\theta,\\
& {e_{r}}^{y}=\cos\frac{\theta}{2}\sin\theta,\\
& {e_{r}}^{z}=\sin\frac{\theta}{2},\\
& {e_{s}}^{x}=-\frac{1}{N}(4R\cos\frac{\theta}{2}+3r\cos\theta+2r)\sin\frac{\theta}{2},\\
& {e_{s}}^{y}=\frac{1}{N}(2R\cos\theta+\frac{1}{2}r\cos\frac{\theta}{2} +\frac{3}{2}r\cos\frac{3\theta}{2}),\\
& {e_{s}}^{z}=\frac{1}{N}r\cos\frac{\theta}{2},\\
& {e_{n}}^{x}=\frac{2}{N}(R\sin\frac{\theta}{2}\cos\theta-r\sin^2\frac{\theta}{2}\sin\theta),\\
& {e_{n}}^{y}=\frac{1}{N}[2R\sin\frac{\theta}{2}\sin\theta+r(\sin^2\theta+\cos\theta)],\\
& {e_{n}}^{z}=-\frac{2}{N}(R\cos\frac{\theta}{2}+r\cos^2\frac{\theta}{2}).
\end{split}
\end{equation}
It is easy to check that ${\rm{U}}_{\mathcal{R}}(r, \theta)$ is a generator of $SU(2)$ group, the universal cover of $SO(3)$, for $\theta\in[0, 4\pi]$.

As another result, the geometry of M\"obius strip will deform its near space denoted as $\Xi\mathbb{M}^2$. The deformation leads to the influence of the normal space on the tangent space, which can be described by a rescaling factor that relates the three-dimensional metric tensor $G_{\alpha\beta}$ and the two-dimensional metric tensor $g_{ab}$. With the definition, $g_{ab}=\partial_a\tb{r}\cdot\partial_b\tb{r}$ $(a, b=1, 2)$, the metric tensor defined on $\mathbb{M}^2$ can be obtained as
\begin{equation}\label{MetricS}
g_{ab}=\left [
\begin{array}{ccc}
 1 & 0\\
 0 & \frac{N^2}{4}
\end{array}
\right ],
\end{equation}
and the corresponding determinant and inverse matrix can be then obtained as
\begin{equation}\label{Detg}
g=\frac{N^2}{4},
\end{equation}
and
\begin{equation}\label{Invg}
g^{ab}=\left [
\begin{array}{ccc}
1 & 0\\
0 & \frac{4}{N^2}
\end{array}
\right ],
\end{equation}
respectively. With the definition, $h_{ab}=\tb{e}_n\cdot\partial^2\tb{r}/\partial q^a\partial q^b$ $(a, b=1, 2)$, the second fundamental form $h_{ab}$ can be written as
\begin{equation}\label{SecFund}
h_{ab}=\left [
\begin{array}{ccc}
0 & R/N\\
R/N & \frac{N^2+r^2}{2N}\sin\frac{\theta}{2}
\end{array}
\right ].
\end{equation}
Further, the Weingarten curvature matrix can be calculated as
\begin{equation}\label{WeinMatrix}
{\alpha_a}^{b}=\left [
\begin{array}{ccc}
0 & -4R/N^3\\
-R/N & -\frac{2(N^2+r^2)}{N^3}\sin\frac{\theta}{2}
\end{array}
\right ]
\end{equation}
with ${\alpha_a}^{b}=-h_{ac}g^{cb}$.

In terms of $\mathbb{M}^2$, the position vector of a point in $\Xi\mathbb{M}^2$ can be parameterized by
\begin{equation}\label{NMobius}
\tb{R}(r, \theta, q_3)=\tb{r}(r, \theta)+q_3\tb{e}_n,
\end{equation}
where $q_3$ is a coordinate variable normal to $\mathbb{M}^2$. With the definition, $G_{\alpha\beta}=\partial_{\alpha}\tb{R}\cdot\partial_{\beta}\tb{R}$  $(\alpha, \beta=1, 2, 3)$, the covariant elements $G_{ab}$ can be described by
\begin{equation}\label{VMetric}
G_{ab}=g_{ab}+(\alpha g+g^T\alpha^T)_{ab}q_3+(\alpha g\alpha^T)_{ab}(q_3)^2
\end{equation}
through $g_{ab}$ and ${\alpha_a}^b$, and $G_{33}=1$ and the rest elements vanishing. It is easy to prove that $g$, the determinant of $g_{ab}$, and $G$, the determinant of $G_{\alpha\beta}$, satisfy the following simple relationship
\begin{equation}\label{Ggrelation}
G=f^2g,
\end{equation}
where $f$ is the rescaling factor with the following form
\begin{equation}\label{Factor}
f=1+{\rm{Tr}}(\alpha)q_3+{\rm{det}}(\alpha)(q_3)^2.
\end{equation}
In relativistic case, the Dirac equation contains only one-order derivative operator, therefore $f^{\frac{1}{2}}$ and $f^{-\frac{1}{2}}$ can be further approximated as
\begin{equation}\label{FactorHalf}
f^{\mp\frac{1}{2}}\approx 1\mp\frac{1}{2}{\rm{Tr}(\alpha)}q_3.
\end{equation}
Obviously, the determinant of ${\alpha_a}^b$ disappears from Eq.~\eqref{FactorHalf}. It means that the geometric effects of rescaling factor depend only on the mean curvature ${\rm{M}}$ of $\mathbb{M}^2$, not on the Gaussian curvature ${\rm{K}}$. ${\rm{M}}$ and ${\rm{K}}$ can be expressed in the following form
\begin{equation}\label{MGCurvature}
\begin{split}
& {\rm{M}}=\frac{1}{2}{\rm{Tr}}(\alpha) =\frac{(N^2+r^2)\sin\frac{\theta}{2}}{N^3},\\
& {\rm{K}}={\rm{det}}(\alpha)=-\frac{4R^2}{N^4},
\end{split}
\end{equation}
respectively. However, in the non-relativistic limit the effective Hamiltonian will contain two-order derivative operators, the factor $f^{\frac{1}{2}}$ and $f^{-\frac{1}{2}}$ have to contain the terms of $(q_3)^2$, and ${\rm{K}}$ will then contribute additional geometric effects. As a result, in the case of a curved surface with nonvanishing ${\rm{K}}$, the thin-layer scheme does not commutate with the non-relativistic limit.

\section{The effective Dirac equation on M\"{o}bius strip}
In this section, for the relativistic particle confined to $\mathbb{M}^2$, the effective Dirac equation will be deduced in the thin-layer quantization formalism. In the semiclassical formalism, a confining potential is first introduced to reduce the degree of freedom normal to $\mathbb{M}^2$. Without loss of generality, the strength of confining potential can not be strong enough to create particle and anti-particle pairs~\cite{Jensen1993Fermions}. In the other words, the number of particles is conserved in the quantization procedure.

\subsection{A Rescaling Transformation}
In quantum mechanics, the particle number conservation can be described by
\begin{equation}\label{ProbConserve}
\begin{split}
\int|\Psi^{\dag}\Psi|d\tau &=\int|\Psi|^2\sqrt{G}d^3q\\
& =\int|\sqrt{f}\Psi|^2(\sqrt{g}drds)dq_3,
\end{split}
\end{equation}
where $\Psi$ is a wave function that describes a relativistic particle in three-dimensional space, $q$ stands for the three curvilinear coordinate variables of an adapted frame, $G$ is the determinant of the metric tensor $G_{\alpha\beta}$ defined in the subspace $\Xi\mathbb{M}^2$ and $g$ is the determinant of the metric tensor $g_{ab}$ defined on the curved surface $\mathbb{M}^2$. The final aim of the thin-layer quantization scheme is to give the effective Dirac equation that is separated from the normal quantum motion analytically. It is worthwhile to notice that the rescaling factor $f$ is generally a function of $q_1$, $q_2$ and $q_3$. The $q$-dependence is determined by the diffeomorphism transformation induced by $\mathbb{M}^2$. In order to divide $\sqrt{G}d^3q$ into a $q_3$-independent part and a $q_{1,2}$-independent one analytically, a new wave function $\chi$, $\chi=\sqrt{f}\Psi$, has to be introduced. The replacement can eliminate $f$ from $\sqrt{G}d^3q$.

Under the diffeomorphism transformation, the wave function $\Psi$ and an ordinary physical operator $\hat{{\rm{O}}}$ satisfy the following transformations
\begin{equation}\label{MResTrans}
\begin{split}
& \chi=f^{\frac{1}{2}}\Psi,\\
& \hat{{\rm{O}}}^{\prime}= f^{\frac{1}{2}}\hat{{\rm{O}}}f^{-\frac{1}{2}}.
\end{split}
\end{equation}
In view of the original intention of the introduction of $\chi$, the above transformation can describe the redistribution of the spatial probability of particle near $\mathbb{M}^2$. However, this transformation can not well describe the influences on spinor. As a result, for the spinor on $\mathbb{M}^2$, one has to consider the rotation transformation defined by the background dreibein field of $\mathbb{M}^2$.

\subsection{A Frame Rotation Transformation}
In comparison with the particle described by Sch\"odinger equation, the particle described by Dirac equation has an additional intrinsic degree of freedom, spin. Since the spinor is not a representation of the group $GL(3, \mathbb{R})$, the diffeomorphism transformation can not describe the dynamics of spinor on $\mathbb{M}^2$. As usual, in $\Xi\mathbb{M}^2$ the spinor is taken as the eigenstate of $\sigma_3$, where $\sigma_3$ is a Pauli matrix. It is straightforward to check that the basis vectors $(\tb{e}_r, \tb{e}_s, \tb{e}_n)$ can be obtained by $(\tb{e}_x, \tb{e}_y, \tb{e}_z)$ through a rotation transformation ${\rm{U}}_{\mathcal{R}}$, which connects the local frames of the different points of $\mathbb{M}^2$. Specifically, $\sigma_3$ can be also obtained from $\sigma_z$ by performing ${\rm{U}}_{\mathcal{R}}$. In the usual Pauli-Dirac representation, for the spinor on $\mathbb{M}^2$ the new wave function $\chi$ and the rescaled physical operator $\hat{O}^{\prime}$ satisfy the following transformations
\begin{equation}\label{FrameTrans}
\begin{split}
& \chi^{\prime}={\rm{U}}_{\mathcal{R}}\chi,\\
& \hat{O}^{\prime\prime}={\rm{U}}_{\mathcal{R}}\hat{O}^{\prime}{\rm{U}}_{\mathcal{R}}^{-1},
\end{split}
\end{equation}
where ${\rm{U}}_{\mathcal{R}}$ denotes the rotation connecting $(\tb{e}_r, \tb{e}_s, \tb{e}_n)$ and $(\tb{e}_x, \tb{e}_y, \tb{e}_z)$, and ${\rm{U}}_{\mathcal{R}}^{-1}$ is the inverse of ${\rm{U}}_{\mathcal{R}}$.

As a conclusion, for a spinor on $\mathbb{S}^2$, the wave function $\Psi$ and an ordinary physical operator $\hat{{\rm{O}}}$ in general need to satisfy the following transformations,
\begin{equation}\label{Trans}
\begin{split}
& \Psi^{\prime}={\rm{U}}_{\mathcal{R}}f^{\frac{1}{2}}\Psi,\\
& \hat{{\rm{O}}}^{\prime}={\rm{U}}_{\mathcal{R}}f^{\frac{1}{2}}\hat{{\rm{O}}}f^{-\frac{1}{2}} {\rm{U}}_{\mathcal{R}}^{-1}.
\end{split}
\end{equation}
Furthermore, the effective physical operator describing the spinor on $\mathbb{S}^2$ can be determined by
\begin{equation}\label{KeyEq}
\hat{{\rm{O}}}_{{\rm{eff}}}=\lim_{q_3\to 0}({\rm{U}}_{\mathcal{R}}f^{\frac{1}{2}}\hat{{\rm{O}}}f^{-\frac{1}{2}} {\rm{U}}_{\mathcal{R}}^{-1})-\hat{{\rm{O}}}_{\bot},
\end{equation}
where $\hat{{\rm{O}}}_{\bot}$ denotes the normal component of $\hat{\rm{O}}$. This equation is key in the present paper, which condenses the initial spirit of the thin-layer quantization formalism~\cite{Wang2016Quantum}. Interestingly, it is can be extended to vector fields, such as electromagnetic field~\cite{Lai2019Geometrical}.

\subsection{Effective Dirac Equation}
In the spirit of the thin-layer quantization scheme, a relativistic particle is initially considered that is described by a Dirac equation,
\begin{equation}\label{DiracEq0}
(i\gamma^{\mu}\partial_{\mu}-m)\Psi=0,
\end{equation}
where $\partial_{\mu}$ is a derivative operator in (3+1)-dimensional spacetime, $m$ is the static mass of particle, and $\gamma^{\mu}$ $(\mu=0, 1, 2, 3)$ are $4\times 4$ Dirac matrices that satisfy the following anticommutation relationship
\begin{equation}\label{DiracM}
[\gamma^{\mu},\gamma^{\nu}]_{+}=\gamma^{\mu}\gamma^{\nu}+\gamma^{\nu}\gamma^{\mu} =2\eta^{\mu\nu},
\end{equation}
where $\eta^{\mu\nu}={\rm diag} (1,-1,-1,-1)$. In Eq.~\eqref{DiracEq0}, $\Psi$ is a four-component spinor, which can be separated into a scalar component $\psi$ and a vector component $\hat{s}$ under the transformations Eqs.~\eqref{MResTrans} and ~\eqref{FrameTrans}. In the Pauli-Dirac representation, the Dirac matrices $\gamma^{\mu}$ can be directly expressed as
\begin{equation}
\gamma^0=\left (
\begin{array}{ccc}
I & 0\\
0 & I
\end{array}
\right ),\quad
\gamma^i=\left (
\begin{array}{ccc}
0 & \sigma^i\\
-\sigma^i & 0
\end{array}
\right ),
\end{equation}
where $I$ is a unit $2\times 2$ matrix, and $\sigma^i$ $(i=x, y, z)$ are Pauli matrices
\begin{equation}
\sigma^x=\left (
\begin{array}{ccc}
0 & 1\\
1 & 0
\end{array}
\right ),
\sigma^y=\left (
\begin{array}{ccc}
0 & -i\\
i & 0
\end{array}
\right ), \sigma^z=\left (
\begin{array}{ccc}
1 & 0\\
0 & -1
\end{array}
\right ),
\end{equation}
respectively.

In the thin-layer quantization formalism, for a relativistic particle in $\Xi\mathbb{M}^2$ the four-component wave function $\Psi$ and the Dirac Hamiltonian $\rm{H}$ in Eq.~\eqref{DiracEq0} should satisfy the transformations Eq.~\eqref{Trans} as
\begin{equation}\label{TransI}
\begin{split}
& \chi={\rm{U}}_{\mathcal{R}}f^{\frac{1}{2}}\Psi,\\
& {\rm{H}}^{\prime}={\rm{U}}_{\mathcal{R}}f^{\frac{1}{2}}{\rm{H}}f^{-\frac{1}{2}} {\rm{U}}_{\mathcal{R}}^{-1}.
\end{split}
\end{equation}
Subsequently, a confining potential is introduced to reduce the normal degree of freedom. In the presence of introduced potential the relativistic particle is going to longtime stay at the ground state of the dimension normal to $\mathbb{M}^2$. With the normal ground state and Eq.~\eqref{KeyEq}, the effective Dirac Hamiltonian can be specifically determined by
\begin{equation}\label{KeyEqMy}
\begin{split}
{\rm{H}}_{{\rm{eff}}}& =\lim_{\varepsilon\to 0}\langle\chi_{{\bot}_0}|({\rm{U}}_{\mathcal{R}}f^{\frac{1}{2}}{\rm{H}}f^{-\frac{1}{2}} {\rm{U}}_{\mathcal{R}}^{-1})-{\rm{H}}_{\bot}|\chi_{{\bot}_0}\rangle\\
&=\langle\chi_{{\bot}_0}|({\rm{U}}_{\mathcal{R}}f^{\frac{1}{2}}{\rm{H}}f^{-\frac{1}{2}} {\rm{U}}_{\mathcal{R}}^{-1})-{\rm{H}}_{\bot}|\chi_{{\bot}_0}\rangle_0,
\end{split}
\end{equation}
where $\varepsilon$ describes the scaling size of the normal degree of freedom, and $\chi_{\bot_0}$ is the normal ground state. In order to obtain the solution of ground state, the transformed wave function $\chi$ is necessarily divided into a tangent part $\chi_{||}(r, s)$ and a normal part $\chi_{\perp}(q_3)$ analytically,
\begin{equation}\label{ChiTN}
\chi(q)=\chi_{\bot}(q_3)\chi_{\|}(r, s).
\end{equation}
In the process, the time dimension is conveniently taken as a constant variable, the introduced confining potential can not strong enough to create particle and antiparticle pairs~\cite{Jensen1993Fermions} to conserve the particle number invariance. For the sake of simplicity, the confining potential $V_c$ can be chosen~\cite{Encinosa2005Wave} in the following form
\begin{equation}\label{ConfPoten0}
V_c=\lim_{\varepsilon\to 0}
\begin{cases}
 0, \quad -\frac{\varepsilon}{2}\leq q_3\leq \frac{\varepsilon}{2},\\
\infty, \quad q_3<-\frac{\varepsilon}{2}, \quad q_3>\frac{\varepsilon}{2},
\end{cases}
\end{equation}
where $\varepsilon$ describes the width of the potential well. And the Dirac equation~\eqref{DiracEq0} is then rewritten as
\begin{equation}\label{Dirac1}
(i\gamma^{\alpha}\partial_{\alpha}-m+V_c)\Psi=E\Psi,
\end{equation}
where $\alpha=1, 2, 3$ describing the tree coordinate variables in an adapted curvilinear coordinate system, and $E$ is the eigenenergy of Dirac particle.

For the extreme limit of $V_c$, the normal component of Dirac equation can be directly written into
\begin{equation}\label{NEq0}
(i\gamma^3\partial_3+V_c)\chi_{\bot} =E_{\bot}\chi_{\bot},
\end{equation}
where $E_{\bot}$ is the normal component of energy that satisfies $E_{\|}+E_{\bot}=E$, wherein $E_{\|}$ is the tangent component. In the presence of $V_c$, the Dirac particle is strictly confined in the interval $[-\varepsilon/2, \varepsilon/2]$, and thus Eq.~\eqref{NEq0} can be simplified as
\begin{equation}\label{NDE0}
i\gamma^3\partial_3\chi_{\bot} =E_{\bot}\chi_{\bot}
\end{equation}
in the interval. Without considering the spin degree of freedom, the above equation can be further rewritten as
\begin{equation}\label{NDE}
-\partial_3^2\chi_{\bot}=E_{\bot}^2\chi_{\bot}.
\end{equation}
With the boundary continuous conditions, the solution of $\chi_{\bot}$ can be given as
\begin{equation}\label{ChiN}
\chi_{\bot}(q_3)=\sqrt{\frac{2}{\varepsilon}}\cos(k_nq_3),
\end{equation}
where $k_n$ is the normal component of momentum that is quantized as $k_n=(2n+1)\pi/\varepsilon$. With the limit $\varepsilon\to 0$, the gap between the ground state and the first excited state becomes enough large, the Dirac particle will permanently lie in the ground state $|\chi_{\bot 0}\rangle$.

In the Pauli-Dirac representation, the spinor on $\mathbb{M}^2$ can be taken as the eigenstates of $\sigma_3$, where $\sigma_3$ is a Pauli matrix in a curvilinear coordinate system. Based on the previously discussions, the new Dirac matrices $\gamma^{\alpha}$ and the new derivative operator $D_{\alpha}$ can be described by the Dirac matrices $\gamma^{i}$ and the derivative operator $\partial_{i}$ in flat space with a rotation transformation $\rm{U}_{\mathcal{R}}$,
\begin{equation}\label{FrameToFrame}
\begin{split}
& \gamma^{\alpha}=\rm{U}_{\mathcal{R}}\gamma^{i}={e^{\alpha}}_{i}\gamma^{i},\\
& D_{\alpha}=\rm{U}_{\mathcal{R}}^{-1}\partial_{i}+\partial_{i}(\rm{U}_{\mathcal{R}}^{-1}) ={e_{\alpha}}^{i}\partial_{i} +(\partial_{i}{e^{i}}_{\alpha}).
\end{split}
\end{equation}
Practically, the rotation transformation $\rm{U}_{\mathcal{R}}$ can be accomplished by performing three rotation transformations ${\rm{U}}_z$, ${\rm{U}}_y$ and ${\rm{U}}_{x}$ in order, where ${\rm{U}}_z$ is a rotation around $\tb{e}_z$, ${\rm{U}}_{y}$ is a rotation around $\tb{e}_y^{\prime}$ which stands for the $y$-axis rotated by ${\rm{U}}_z$, and ${\rm{U}}_x$ stands for a rotation around $\tb{e}_x^{\prime\prime}$ which stands for the $x$-axis rotated by performing ${\rm{U}}_z$ and then ${\rm{U}}_y$. In Cartesian coordinate system, ${\rm{U}}_z$ can be deduced as
\begin{equation}\label{Rot-z}
{\rm{U}}_z(\theta)=\left [
\begin{array}{ccc}
\cos\theta & \sin\theta & 0\\
-\sin\theta & \cos\theta & 0\\
0 & 0 & 1
\end{array}
\right ],
\end{equation}
${\rm{U}}_{y}$ can be
\begin{equation}\label{Rot-y}
{\rm{U}}_{y}(\theta)=\left [
\begin{array}{ccc}
\cos\frac{\theta}{2} & 0 & -\sin\frac{\theta}{2}\\
0 & 1 & 0\\
\sin\frac{\theta}{2} & 0 & \cos\frac{\theta}{2}
\end{array}
\right ],
\end{equation}
and ${\rm{U}}_x$ can be
\begin{equation}\label{Rot-x}
{\rm{U}}_{x}(r,\theta)=\left [
\begin{array}{ccc}
1 & 0 & 0\\
0 & \cos\theta_x & \sin\theta_x\\
0 & -\sin\theta_x & \cos\theta_x
\end{array}
\right ],
\end{equation}
respectively, here $\cos\theta_x=2(R+r\cos\frac{\theta}{2})/N$ and $\sin\theta_x=r/N$. It is easy to check that ${\rm{U}}_{\mathcal{R}}={\rm{U}}_x{\rm{U}}_y{\rm{U}}_z$. In the Pauli-Dirac presentation of $\sigma_3$, ${\rm{U}}_{\mathcal{R}}$ can be reexpressed as
\begin{equation}\label{RotTrans}
{\rm{U}}_{\mathcal{R}}(r, \theta)=e^{-i{\bm{\theta}}\cdot{\bm{\sigma}}},
\end{equation}
where ${\bm\theta}=(\theta_x, \theta_y, \theta_z)$ and ${\bm\sigma}=\bm{e}_n\sigma_3$, with $\theta_x={\rm{arcsin}}(r/N)$, $\theta_y=\theta/2$, $\theta_z=\theta$.

Substitute Eqs.~\eqref{FactorHalf},~\eqref{ChiN} and~\eqref{RotTrans} into Eq.~\eqref{KeyEqMy}, we can obtain the effective Hamiltonian $\rm{H}_{eff}$ that describes the relativistic particle on $\mathbb{M}^2$ as
\begin{equation}\label{EffHam}
{\rm{H}_{eff}}=i\gamma^{a}(\partial_a-i\sigma_3\mathcal{A}_a)-(m+m_{\rm eff}),
\end{equation}
where $\mathcal{A}$ is a geometric gauge potential with $\mathcal{A}_{a}=\partial_{a}({\bm{\theta}}\cdot{\bm{\sigma}})$ $(a=r, s)$ and $m_{\rm eff}$ is an effective mass induced by geometry, $m_{\rm eff}=\frac{1}{2}{\rm{Tr}}(\alpha)$, wherein $\alpha$ denotes the Weingartein curvature matrix. In terms of Eq.~\eqref{EffHam}, the effective Dirac equation can be written into the following form
\begin{equation}\label{EffDiracEq}
[i\gamma^a(\partial_a-i\sigma_3\mathcal{A}_a)-(m+m_{\rm eff})]|\chi_{\|}\rangle=E_{\|}|\chi_{\|}\rangle,
\end{equation}
where $E_{\|}$ is the tangent component of energy eigenvalue, and $|\chi_{\|}\rangle$ stands for the tangent part of wave function. Strikingly, the geometry of $\mathbb{M}^2$ plays a role of gauge potential and the mean curvature of $\mathbb{M}^2$ contributes an effective mass.

Obviously, the geometric gauge potential $\mathcal{A}_{a}$ is given by $\partial_{a}$ acting on ${\rm{U}_{\mathcal{R}}}^{-1}$, and the effective mass $m_{\rm{eff}}$ results from the action of $\partial_3$ on $f^{-\frac{1}{2}}$. It is worthwhile to notice that all the high order terms of $q_3$ in $f^{-\frac{1}{2}}$ are vanished by performing the integral $\langle\chi_{\bot_0}|\chi_{\bot_0}\rangle_0$. Meaningly, the high power terms are missing the actions of $\partial_3$ before performing the non-relativistic limit. In other words, the thin-layer quantization scheme does not commute with the non-relativistic limit process. Importantly, the effective Pauli equation on $\mathbb{M}^2$ should be performed in a certain order~\cite{Wang2016Quantum}, specifically, the non-relativistic limit is prior to the thin-layer quantization scheme.

\section{Geometry-induced Monopole Magnetic Field}
\begin{figure}[htbp]
\centering
% Requires \usepackage{graphicx}
(a)\includegraphics[width=0.2\textwidth]{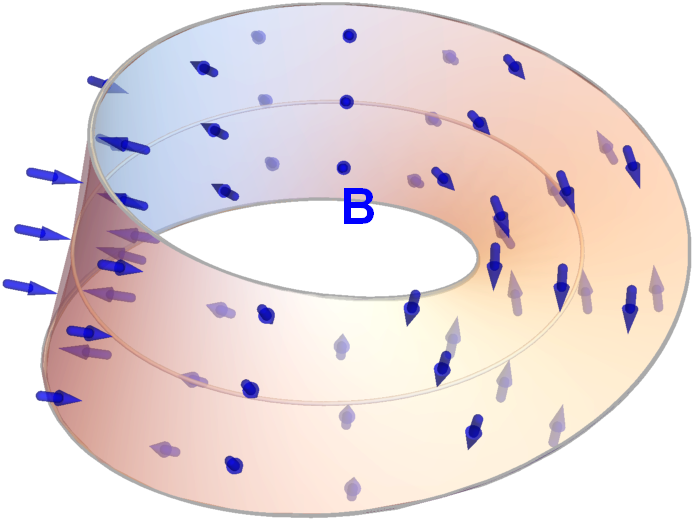}\quad (b)\includegraphics[width=0.21\textwidth]{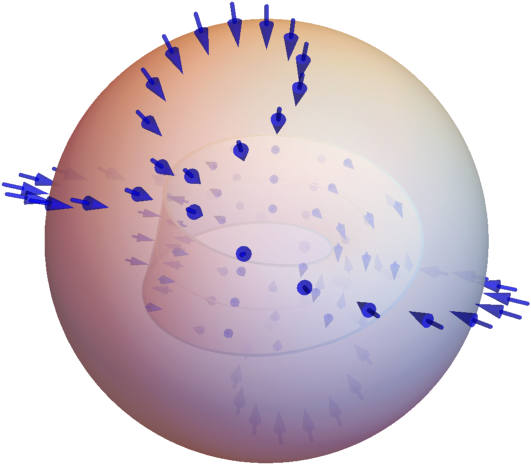}\\
(c)\includegraphics[width=0.2\textwidth]{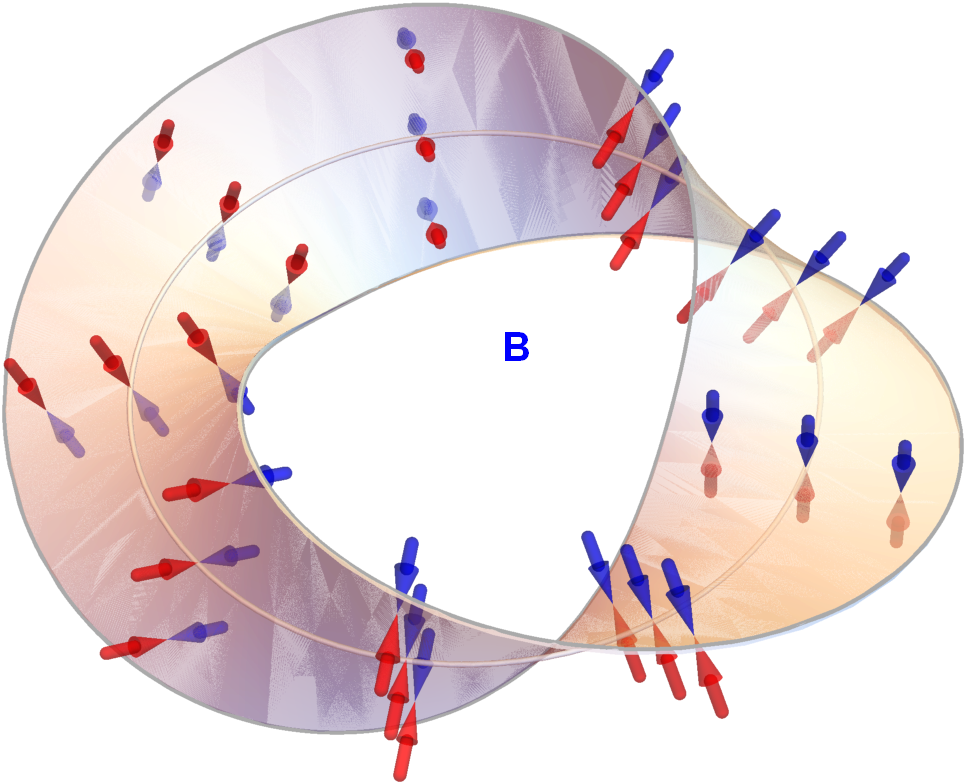}\quad
(d)\includegraphics[width=0.18\textwidth]{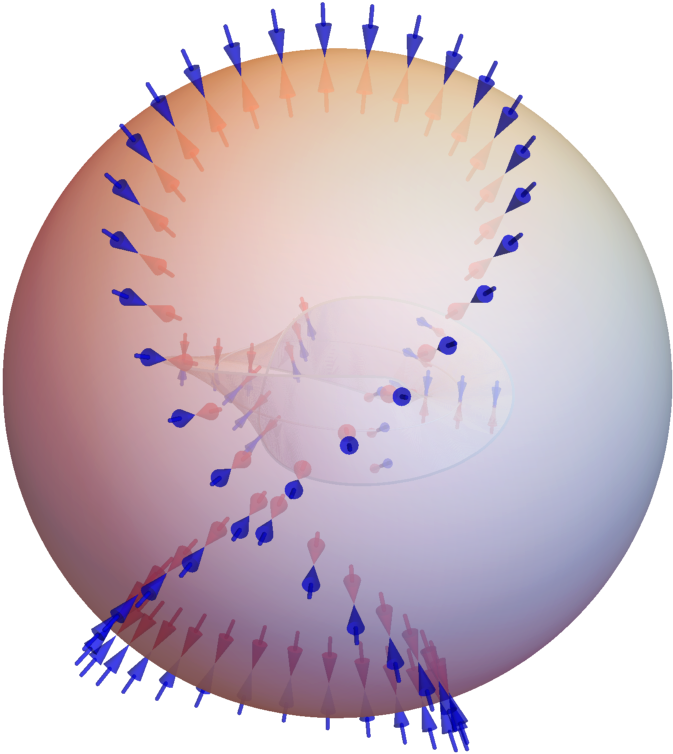}
\caption{\footnotesize (a) In the case of a half-integer linking number, there is an effective monopole magnetic field $\mathcal{\bm{B}}$ that orientates inward. (b) The monopole field is projected onto a  sphere. (c) In the case of a integer linking number, there is an effective common magnetic field. (d) The common field is projected onto a sphere.}\label{Fig2}
\end{figure}
\begin{figure}[htbp]
\centering
% Requires \usepackage{graphicx}
\includegraphics[width=0.35\textwidth]{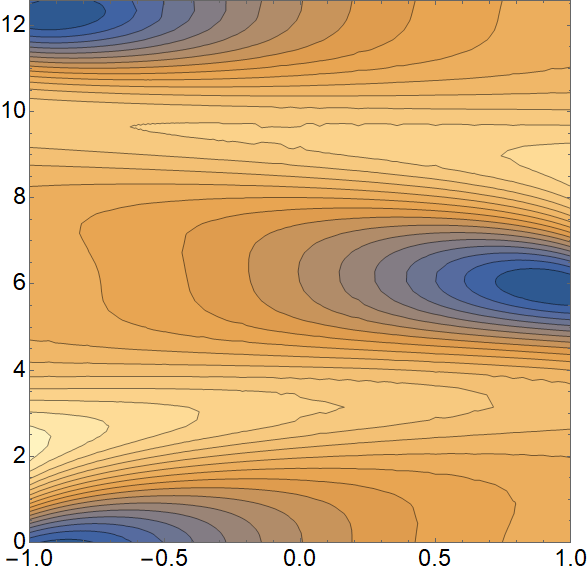}
\includegraphics[width=0.07\textwidth]{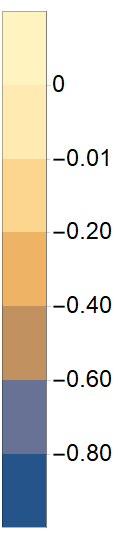}
\caption{\footnotesize The contours of geometric magnetic field in the plant spanned by $r$ and $\theta$ with $R=4$, $r\in[-1,1]$ and $\theta\in[0, 4\pi]$.}\label{Fig3}
\end{figure}

The rotation ${\rm{U}}_{\mathcal{R}}$ describes the connection of the local frames of different points on $\mathbb{M}^2$, which can be described by two tangent coordinate variables $r$ and $s$ of $\mathbb{M}^2$ without the normal coordinate variable $q_3$. It is easy to obtain that the three components of the geometric gauge potential are $\mathcal{A}_r=2R/N^2$,
\begin{widetext}
\begin{equation} \mathcal{A}_s=\frac{1}{\sqrt{N^2-r^2}}\sin^2\frac{\theta}{2}\sin^2\theta_x(\cos\theta\cos^2\theta_x +\cos\frac{\theta}{2}\sin 2\theta_x)
 +\frac{1}{N}[(\sin\theta\sin\frac{\theta}{2}+2\cos\frac{\theta}{2})\cos\theta_x -\cos\theta\sin\theta_x],
\end{equation}
\end{widetext}
and $\mathcal{A}_n=0$, respectively. And the geometric magnetic field can be then calculated as
\begin{equation}\label{GeometricMF}
\mathcal{B}_n =\partial_r\mathcal{A}_{s}-\partial_s\mathcal{A}_r,
\end{equation}
and $\mathcal{B}_r=\mathcal{B}_s=0$, because $\mathcal{A}_r$ and $\mathcal{A}_s$ both do not depend on $q_3$, and $\mathcal{A}_3=0$. It is apparent that $\mathcal{B}_n$ is along the normal direction of $\mathbb{M}^2$, and orientates inward that is sketched in Fig.~\ref{Fig2} (a) as an effective monopole magnetic field. The result is described in Fig.~\ref{Fig2} (b). The particular phenomenon just displays in the case of a half-integer linking number~\cite{Korte2009Curvature} $\frac{1}{2\pi}\int{\tau d\theta}=n+\frac{1}{2}$ $(n\in Z)$, $Z$ is an integer. For an integer linking number $\frac{1}{2\pi}\int{\tau d\theta}=n+1$, the effective magnetic field becomes common that is sketched in Fig.~\ref{Fig2}(c), which is also projected onto a sphere described in Fig.~\ref{Fig2} (d). It is apparently that the monopole magnetic field is completely determined by the topological structure of $\mathbb{M}^2$, a single side. And the geometric magnetic field $\mathcal{B}_n$ is specifically described in the plane spanned by $r$ and $\theta$ as sketched in Fig.~\ref{Fig3}.

Noticeably, the geometric magnetic field is distinctly different from the common magnetic field. The geometric magnetic field acts on a particle by coupling with spin, while the common magnetic field acts on a particle by coupling with electric charge. Specifically, the effective magnetic field is determined by the geometry of $\mathbb{M}^2$, the nontrivial monopole properties are defined by the nontrivial topological properties of $\mathbb{M}^2$. Once the nontrivial single side vanishes, the nontrivial monopole would disappear and the common effective magnetic field would appear. Furthermore, the geometric magnetic field is in general a non-Abelian gauge field, which determines the gauge structure of the effective Hamiltonian confined on $\mathbb{M}^2$. In other words, the gauge structure of effective dynamics can be constructed by the geometry of a particular curved surface. As potential applications, the effective gauge field can be generated by designing the geometries two-dimensional nanodevices.

\section{Geometry-induced Quantum Spin Hall Effect}

\begin{figure}[htbp]
\centering
% Requires \usepackage{graphicx}
\includegraphics[width=0.4\textwidth]{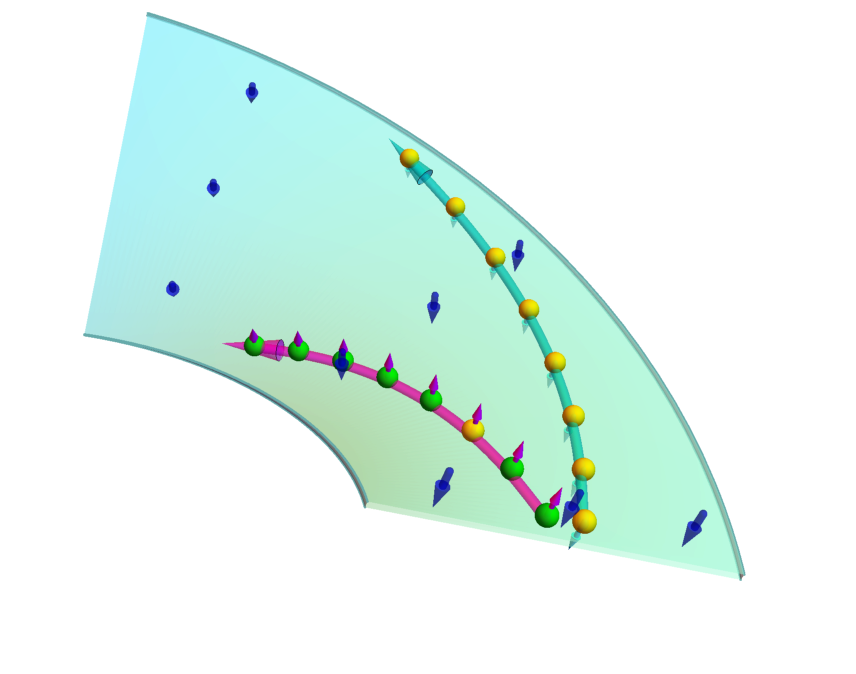}
\caption{\footnotesize Schematic of a spin Hall effect on a part of M\"obius strip. The blue arrow denotes the geometric magnetic field, the green balls with pink upward arrows are the electrons of spin up, the green balls with cyan downward arrows are the electrons of spin down, and the the red arrows stand for the moving direction of electrons.}\label{Fig4}
\end{figure}
\begin{figure}[htbp]
\centering
% Requires \usepackage{graphicx}
\includegraphics[width=0.4\textwidth]{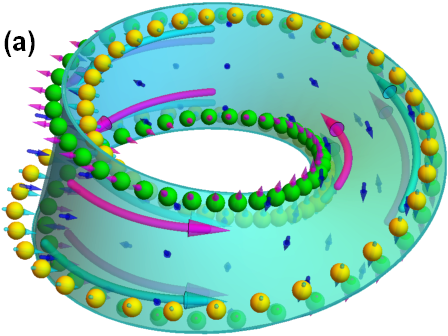}\\
\smallskip\smallskip
\includegraphics[width=0.4\textwidth]{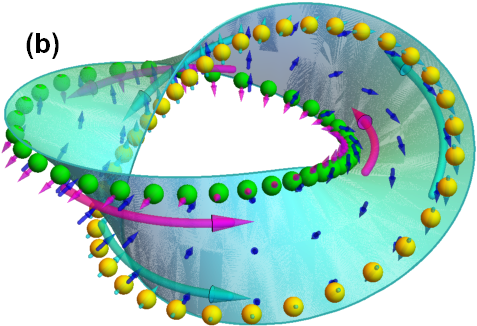}
\caption{\footnotesize (a) Spin Hall effect on the M\"obius strip of a half-integer linking number $\frac{1}{2}$. (b) Spin Hall effect on the M\"obius strip of an integer linking number $1$. The blue arrows stand for the geometric magnetic field, the green balls with pink outward arrows denote the spin-outward particles, the yellow balls with cyan inward arrows are the spin-inward particles. The big pink arrows stand for the current consisting of spin-outward particles, and the big cyan arrows stand for the current consisting of the spin-inward ones.}\label{Fig5}
\end{figure}

In the presence of $\mathcal{\bm{A}}$, the Dirac particle moving on $\mathbb{M}^2$ will then feel a pseudo-Lorentz force, which is induced by the geometry of $\mathbb{M}^2$. Because that the spin of particle is initially coupled with the geometry of $\mathbb{M}^2$ as the term $\sigma_{3}\mathcal{A}$ in Eq.~\eqref{EffDiracEq}~\cite{Wang2020Geometry}, which plays the role of the spin-orbit coupling~\cite{Kane2005Quantum} in conventional semiconductor and that in the presence of a strain gradient~\cite{Zhang2006Quantum}, and which can be rewritten as $u_3\mathcal{B}_n$ in the non-relativistic limit, where $u_3$ denotes the normal component of spin magnetic moment. In light of the left hand rule, for a certain section of $\mathbb{M}^2$ the spin-outward particles gather to one side, the spin-inward particles aggregate toward the other side, which are sketched in Fig.~\ref{Fig4}. The spin-outward means that the spin orientates outward along the normal direction of $\mathbb{M}^2$, the spin-inward means that the spin orientates inward $\mathbb{M}^2$. As a result, on $\mathbb{M}^2$ the spin-outward particles and the spin-inward particles are completely separated into two groups, they are gathering on the two sides for a certain section, respectively, and they are collecting on two different faces for a certain edge, respectively. Most strikingly, the spin-inward particles and the spin-outward particles are moving in a same direction with same spin polarization as a full spin current for a certain edge that sketched in Fig.~\ref{Fig5} (a). Interestingly, the spin polarization is entirely determined by the nontrivial topological properties of $\mathbb{M}^2$, which determines the degeneracy of pseudo-Landau levels in momentum space. The two degenerate pseudo-Landau levels are separated in configuration space with a gap that is about the width of $\mathbb{M}^2$. Mathematically, the M\"obius strip is a two-dimensional compact manifold with a single boundary and a one-sided surface. In other words, the topological structure of $\mathbb{M}^2$ can spontaneously flip the "spin-down" into the "spin-up" to provide a pure spin current. For a half-integer number  $\frac{1}{2\pi}\oint\tau ds=n+\frac{1}{2}$ $(n\in\mathbb{Z})$, the pure spin current commonly attributes to the spin-outward particles and the spin-inward particles. In the case of $\frac{1}{2}$, the big cyan arrows and the big pink ones contribute equally to the spin current along the same edge. In the case of an integer number $\frac{1}{2\pi}\oint\tau ds=n$, the pure spin current differently attributes one kind of particle contribution, either the spin-outward particles or the spin-inward ones. The result is sketched in Fig.~\ref{Fig5}(b). Distinguishingly, the spin-outward particles and the spin-inward ones do not have the same contribution to the spin current, and they are impossibly along the same edge.

As a classical analogy, this can be thought of in terms of the magnus effect, a spinning soccer ball will "stray" from its normal straight path in a direction dependent on it's sense of rotation. Therefore, the spin-outward particles will initially gather toward one side, the spin-inward ones aggregate toward the other side for a certain section of $\mathbb{M}^2$. In the case of half-integer linking number, the "so-called" two edges are really the same one, the emergence of quantum spin Hall effect is eventually determined by the geometry of $\mathbb{M}^2$. In the case of integer linking number, the two edges are completely different, the spin current vanishes for the whole of $\mathbb{M}^2$, the spin-outward current emerges along one edge, while the spin-inward current does along the other.

\section{Conclusions and discussions}
For the quantum particles confined to a two-dimensional curved system, the geometric effects are a standing-long interesting topic. The required effective dynamics can be given by using the thin-layer quantization scheme, which is suitable and valid. Because that two important geometric effects, the geometric potential and the geometric momentum, have been proved by experiments. The two effects both result from the rescaling factor that depend on the normal coordinate variable. The dependence of normal space directly originates from the metric tensor that is defined in the three-dimensional subspace spanned by two tangent coordinate variables and a normal one of $\mathbb{S}^2$. So far, the fundamental framework of thin-layer quantization scheme is suitable and sufficient for the geometric potential and geometric momentum~\cite{Costa1981Quantum, Liu2011Geometric, Wang2017Geometric}. Unfortunately, the fundamental formalism can not give the geometric gauge potential that is related to the symmetries of the introduced confining potential~\cite{Wang2018Geometric} for the curved surface embedded in the usual three-dimensional Euclidean space~\cite{Wang2020Geometry}. In order to remove the difficulty, the formula of geometric effects are rediscussed, which is clearly evidenced by that the geometric effects are not only from the rescaling transformation, but also from the rotation transformation intimately connected to the local frames. These results will enable the thin-layer quantization formalism to play a more important and effective role in the effective quantum dynamics for the particles confined to low-dimensional curved systems, especially with the development of low-dimensional nanodevices.

For the particles confined to a M\"obius strip, the effective Dirac equation is given by using the developed thin-layer quantization formalism. There are two important results for the geometric effects. One is the effective mass that results from the rescaling factor. The other is the effective gauge potential that results from the rotation transformation connected to the local frames. The presence of the rescaling factor determines that the thin-layer quantization formalism does not commutate with the non-relativistic limit. Interestingly, the effective magnetic field is monopole that is determined by the single face of M\"obius strip. This result provides a feasible way to generate a non-Abelian monopole magnetic field, the gauge structure can be constructed by designing the geometry of two-dimensional curved system. In the presence of monopole effective magnetic field, the spin-outward particles and the spin-inward ones are completely separated as full spin polarization. The pure spin current is also determined by the geometry of M\"obius strip due to the coupling of geometry and spin. As a conclusion, the nontrivial topological properties of M\"obius strip entirely determine the emergences of the monopole magnetic field and the quantum spin Hall effect. In other words, the complex geometries and topologies of two-dimensional systems can provide a new perspective to investigate new phenomena of Hall physics implied in a high-dimensional space.

With  the rapid development of flexible electronics, flexible spintronics and metastucture physics, the geometric quantum effects play a more and more important role in the effective quantum dynamics for two-dimensional curved systems. The geometries of nanodevices can be employed to improve the development of nanodevices and topological quantum computation and topological quantum commutation. Altogether, our results demonstrate a viable manner to control the electronic levels and transpose properties of two-dimensional curved system, shedding new light on the design of novel electronics devices by geometry engineering.

\section*{Acknowledgments}
This work is jointly supported by the Natural Science Foundation of Shandong Province of China (Grant No. ZR2020MA091), the National Key R\&D Program of China (Grant No. 2017YFA0303702), the National Nature Science Foundation of China (Grants, No. 11625418).

\bibliographystyle{apsrev4-1}
\bibliography{DiracMobius}

\end{document}